  \documentclass[apj]{emulateapj}
  \usepackage{apjfonts}

\shorttitle{MAGNETIC ACTIVITY CYCLE IN $\iota$~HOR}
\shortauthors{METCALFE ET AL.}

\begin{document}
 
\title{Discovery of a 1.6-year Magnetic Activity Cycle in the Exoplanet 
Host Star $\iota$~Horologii}

\author{T.~S.\ Metcalfe\altaffilmark{1}, 
	S.\ Basu\altaffilmark{2}, 
	T.~J.\ Henry\altaffilmark{3}, 
	D.~R.\ Soderblom\altaffilmark{4}, 
	P.~G.\ Judge\altaffilmark{1}, 
	M.\ Kn{\"o}lker\altaffilmark{1},
	S.\ Mathur\altaffilmark{1}, 
	M.\ Rempel\altaffilmark{1}
}

\altaffiltext{1}{High Altitude Observatory, NCAR, P.O.\ Box 3000, Boulder, 
CO 80307}
\altaffiltext{2}{Department of Astronomy, Yale University, P.O.\ Box 
208101, New Haven, CT 06520}
\altaffiltext{3}{Department of Physics and Astronomy, Georgia State 
University, Atlanta, GA 30302}
\altaffiltext{4}{Space Telescope Science Institute, 3700 San Martin Dr., 
Baltimore MD 21218}

\begin{abstract}

The Mount Wilson Ca HK survey revealed magnetic activity variations in a 
large sample of solar-type stars with timescales ranging from 2.5 to 25 
years. This broad range of cycle periods is thought to reflect differences 
in the rotational properties and the depths of the surface convection 
zones for stars with various masses and ages. In 2007 we initiated a 
long-term monitoring campaign of Ca~{\sc ii} H and K emission for a sample 
of 57 southern solar-type stars to measure their magnetic activity cycles 
and their rotational properties when possible. We report the discovery of 
a 1.6-year magnetic activity cycle in the exoplanet host star 
$\iota$~Horologii, and we obtain an estimate of the rotation period that 
is consistent with Hyades membership. This is the shortest activity cycle 
so far measured for a solar-type star, and may be related to the 
short-timescale magnetic variations recently identified in the Sun and 
HD\,49933 from helio- and asteroseismic measurements. Future asteroseismic 
observations of $\iota$~Hor can be compared to those obtained near the 
magnetic minimum in 2006 to search for cycle-induced shifts in the 
oscillation frequencies. If such short activity cycles are common in F 
stars, then NASA's {\it Kepler} mission should observe their effects in 
many of its long-term asteroseismic targets.

\end{abstract}

\keywords{stars: activity---stars: chromospheres---stars: 
individual(HD~17051, HR~810)---surveys}


\section{BACKGROUND}

Astronomers have been making telescopic observations of sunspots since the 
time of Galileo, gradually building a historical record showing a periodic 
rise and fall in the number of sunspots every 11 years. We now know that 
sunspots are regions with a sufficiently strong magnetic field to alter 
the local thermal structure, so this 11-year sunspot cycle actually traces 
a variation in surface magnetism. Attempts to understand this behavior 
theoretically often invoke a combination of differential rotation, 
convection, and meridional flow to modulate the global field through a 
magnetic dynamo \citep[e.g., see][]{rem06}. Although we cannot observe 
spots on other solar-type stars directly these areas of concentrated 
magnetic field produce, among other signatures, strong emission in the 
Ca~{\sc ii}~H (396.8~nm) and K (393.4~nm) spectral lines. The intensity of 
the emission scales with the amount of non-thermal heating in the 
chromosphere, making these lines a useful spectroscopic proxy for the 
strength of, and fractional area covered by, magnetic fields 
\citep{lei59}. \cite{wil78} was the first to demonstrate that many 
solar-type stars exhibit long-term cyclic variations in their Ca~{\sc ii} 
H and K (hereafter Ca~HK) emission, analogous to the solar variations.

Significant progress in dynamo modeling emerged after helioseismology 
provided meaningful constraints on the Sun's interior structure and 
dynamics \citep{bro89,sch98}. Variations in the mean strength of the solar 
magnetic field lead to significant shifts ($\sim$0.5~$\mu$Hz) in the 
frequencies of even the lowest-degree p-modes \citep{lw90,sal04}. 
Space-based asteroseismology missions, such as {\it MOST} \citep{wal03}, 
{\it CoRoT} \citep{bag06}, and {\it Kepler} \citep{bor10}, as well as 
ground-based networks like the Stellar Observations Network Group 
\citep[SONG;][]{gru08}, are now allowing additional tests of dynamo models 
using other solar-type stars \citep[e.g., see][]{cha07,met07}.

The F8V star $\iota$~Horologii ($\iota$~Hor $\equiv$ HD\,17051 $\equiv$ 
HR\,810, V=5.4, B$-$V=0.57) hosts a non-transiting 2~$M_J$ exoplanet with 
an orbital period of 311 days \citep{kur00,nae01}. Although it is 
currently situated in the southern hemisphere, kinematic considerations 
have led to the suggestion that it could be an evaporated member of the 
Hyades cluster \citep{mon01}. Asteroseismic observations support this 
conclusion, since the acoustic oscillation frequencies of the star are 
best reproduced with models that have the same metallicity, helium 
abundance, and stellar age as other Hyades members \citep{vau08}.

We report the discovery of a 1.6-year magnetic activity cycle in 
$\iota$~Hor from synoptic Ca~HK measurements obtained with the Small and 
Moderate Aperture Research Telescope System (SMARTS) 1.5-m telescope at 
Cerro Tololo Interamerican Observatory (CTIO) since 2008. We provide an 
overview of the survey methodology and analysis methods in \S\ref{SEC2}, 
and we present the stellar activity measurements and other derived 
properties in \S\ref{SEC3}. We conclude with a discussion of the broader 
implications of this discovery for stellar dynamo modeling and future 
observations in \S\ref{SEC4}.


  \tablewidth{0pt}
  \tabletypesize{\footnotesize}
  \tablecaption{Journal of observations for $\iota$~Hor.\label{tab1}}
  \begin{deluxetable*}{ccccc|ccccc}
  \tablehead{\colhead{DATE}&\colhead{UT}&\colhead{HJD\,(2450000+)}&\colhead{$S_{\rm MWO}$}&\colhead{$\sigma_S$}&
  \colhead{DATE}&\colhead{UT}&\colhead{HJD\,(2450000+)}&\colhead{$S_{\rm MWO}$}&\colhead{$\sigma_S$}}
  \startdata
  2008-02-15 & 01:18:19 & 4511.55234 & 0.2349 & 0.0023 & 2009-01-31 & 02:22:28 & 4862.59740 & 0.2625 & 0.0018 \\
  2008-02-15 & 01:19:33 & 4511.55319 & 0.2350 & 0.0022 & 2009-02-08 & 01:07:19 & 4870.54491 & 0.2716 & 0.0021 \\
  2008-03-16 & 23:49:03 & 4542.48971 & 0.2282 & 0.0014 & 2009-02-08 & 01:08:33 & 4870.54576 & 0.2676 & 0.0024 \\
  2008-03-16 & 23:50:17 & 4542.49056 & 0.2340 & 0.0014 & 2009-02-24 & 00:59:15 & 4886.53882 & 0.2658 & 0.0019 \\
  2008-07-09 & 09:06:28 & 4656.87999 & 0.2324 & 0.0022 & 2009-02-24 & 01:00:29 & 4886.53967 & 0.2634 & 0.0019 \\
  2008-07-09 & 09:07:42 & 4656.88085 & 0.2333 & 0.0023 & 2009-03-05 & 00:59:01 & 4895.53846 & 0.2647 & 0.0019 \\
  2008-07-25 & 09:32:42 & 4672.89890 & 0.2279 & 0.0016 & 2009-03-05 & 01:00:15 & 4895.53932 & 0.2667 & 0.0019 \\
  2008-07-25 & 09:33:56 & 4672.89975 & 0.2365 & 0.0017 & 2009-03-28 & 23:58:09 & 4919.49599 & 0.2747 & 0.0019 \\
  2008-07-27 & 08:52:52 & 4674.87132 & 0.2178 & 0.0016 & 2009-03-28 & 23:59:22 & 4919.49683 & 0.2721 & 0.0018 \\
  2008-07-27 & 08:54:05 & 4674.87216 & 0.2219 & 0.0017 & 2009-07-01 & 09:52:11 & 5013.91137 & 0.2738 & 0.0021 \\
  2008-08-03 & 10:02:01 & 4681.91961 & 0.2120 & 0.0012 & 2009-07-01 & 09:53:25 & 5013.91222 & 0.2762 & 0.0022 \\
  2008-08-03 & 10:03:15 & 4681.92046 & 0.2198 & 0.0012 & 2009-08-04 & 09:34:58 & 5047.90085 & 0.2563 & 0.0022 \\
  2008-08-18 & 09:32:47 & 4696.89982 & 0.2269 & 0.0014 & 2009-08-04 & 09:36:12 & 5047.90171 & 0.2566 & 0.0023 \\
  2008-08-18 & 09:34:01 & 4696.90067 & 0.2264 & 0.0014 & 2009-09-13 & 06:47:36 & 5087.78568 & 0.2349 & 0.0013 \\
  2008-09-07 & 05:12:15 & 4716.71937 & 0.2221 & 0.0013 & 2009-09-13 & 06:48:50 & 5087.78653 & 0.2326 & 0.0013 \\
  2008-09-07 & 05:13:28 & 4716.72022 & 0.2235 & 0.0013 & 2009-10-17 & 04:40:09 & 5121.69716 & 0.2334 & 0.0015 \\
  2008-09-15 & 06:05:01 & 4724.75613 & 0.2319 & 0.0015 & 2009-10-17 & 04:41:23 & 5121.69802 & 0.2326 & 0.0015 \\
  2008-09-15 & 06:06:15 & 4724.75699 & 0.2331 & 0.0015 & 2009-11-03 & 05:06:12 & 5138.71491 & 0.2232 & 0.0016 \\
  2008-09-25 & 04:50:18 & 4734.70432 & 0.2206 & 0.0012 & 2009-11-03 & 05:07:26 & 5138.71577 & 0.2281 & 0.0017 \\
  2008-09-25 & 04:51:31 & 4734.70516 & 0.2264 & 0.0013 & 2009-11-18 & 03:17:39 & 5153.63906 & 0.2269 & 0.0034 \\
  2008-10-04 & 05:20:31 & 4743.72530 & 0.2276 & 0.0013 & 2009-11-18 & 03:18:53 & 5153.63991 & 0.2202 & 0.0032 \\
  2008-10-04 & 05:21:45 & 4743.72616 & 0.2291 & 0.0014 & 2009-12-18 & 04:21:32 & 5183.68213 & 0.2204 & 0.0015 \\
  2008-10-12 & 05:16:34 & 4751.72251 & 0.2368 & 0.0027 & 2009-12-18 & 04:22:46 & 5183.68299 & 0.2212 & 0.0014 \\
  2008-10-12 & 05:17:48 & 4751.72336 & 0.2320 & 0.0028 & 2009-12-24 & 04:08:00 & 5189.67244 & 0.2131 & 0.0017 \\
  2008-10-25 & 05:56:51 & 4764.75028 & 0.2314 & 0.0015 & 2009-12-24 & 04:09:14 & 5189.67330 & 0.2134 & 0.0015 \\
  2008-10-25 & 05:58:05 & 4764.75114 & 0.2307 & 0.0015 & 2010-01-09 & 02:54:30 & 5205.62063 & 0.2190 & 0.0020 \\
  2008-11-02 & 06:32:48 & 4772.77507 & 0.2244 & 0.0015 & 2010-01-09 & 02:55:44 & 5205.62149 & 0.2112 & 0.0020 \\
  2008-11-02 & 06:34:02 & 4772.77592 & 0.2346 & 0.0016 & 2010-02-26 & 01:21:21 & 5253.55412 & 0.2271 & 0.0016 \\
  2008-11-10 & 05:11:55 & 4780.71867 & 0.2364 & 0.0015 & 2010-02-26 & 01:22:35 & 5253.55498 & 0.2288 & 0.0016 \\
  2008-11-10 & 05:13:09 & 4780.71953 & 0.2361 & 0.0015 & 2010-03-21 & 23:41:44 & 5277.48459 & 0.2260 & 0.0017 \\
  2008-11-26 & 06:04:01 & 4796.75427 & 0.2391 & 0.0015 & 2010-03-21 & 23:42:58 & 5277.48545 & 0.2186 & 0.0016 \\
  2008-11-26 & 06:05:14 & 4796.75510 & 0.2311 & 0.0014 & 2010-07-22 & 10:10:53 & 5399.92526 & 0.2542 & 0.0021 \\
  2008-12-02 & 04:04:26 & 4802.67098 & 0.2341 & 0.0017 & 2010-07-22 & 10:12:07 & 5399.92612 & 0.2488 & 0.0023 \\
  2008-12-02 & 04:05:40 & 4802.67183 & 0.2311 & 0.0016 & 2010-07-31 & 09:42:34 & 5408.90597 & 0.2513 & 0.0034 \\
  2009-01-13 & 02:57:40 & 4844.62263 & 0.2494 & 0.0015 & 2010-07-31 & 09:44:35 & 5408.90737 & 0.2515 & 0.0035 \\
  2009-01-13 & 02:58:54 & 4844.62349 & 0.2480 & 0.0015 & 2010-08-09 & 09:02:53 & 5417.87874 & 0.2389 & 0.0016 \\
  2009-01-31 & 02:21:14 & 4862.59654 & 0.2638 & 0.0018 & 2010-08-09 & 09:04:06 & 5417.87959 & 0.2468 & 0.0023 
  \enddata
  \end{deluxetable*}

\section{OBSERVATIONS \& DATA REDUCTION \label{SEC2}}

The chromospheric activity survey of \cite{hen96} contained a total of 
1016 observations of 815 individual stars with visual magnitudes between 
0.0 and about 9.0, which were observed using the {\it RC Spec} instrument 
on the CTIO 1.5-m telescope. Several sub-samples were defined, including 
the ``Best \& Brightest'' (B) and ``Nearby'' (N) samples, which together 
contain 92 individual stars with visual magnitudes between 0.0 and 7.9, 
and B$-$V colors that are approximately solar. In August 2007, we began a 
long-term Ca~HK monitoring program for the 57 stars in the combined (B+N) 
sample that are brighter than V$=6$ \citep{met09}, the limiting magnitude 
of future ground-based asteroseismic observations by SONG. All of the most 
promising southern asteroseismic targets are included in this B+N 
sub-sample.

Since January 2008, we have obtained 74 low-resolution ($R\sim2500$) 
spectra of $\iota$~Hor covering 37 distinct epochs with the upgraded {\it 
RC Spec} instrument on the SMARTS 1.5-m telescope. Using standard 
IRAF\footnote{IRAF is distributed by the National Optical Astronomy 
Observatory, which is operated by the Association of Universities for 
Research in Astronomy (AURA) under cooperative agreement with the National 
Science Foundation.} routines, the 60\,s integrations were subjected to 
the usual bias and flat field corrections and the spectra were extracted 
and wavelength calibrated using a reference He-Ar spectrum obtained 
immediately before the stellar exposures. Following \cite{dun91}, the 
calibrated spectra were then integrated in 1.09~\AA\ triangular passbands 
centered on the cores of the Ca H and K lines and compared to 20~\AA\ 
continuum passbands from the wings of the lines to generate a CTIO 
chromospheric activity index, $S_{\rm CTIO}$. These values were converted 
to Mount Wilson indices ($S_{\rm MWO}$) using data for 26 targets that 
were observed contemporaneously with the Solar-Stellar Spectrograph at 
Lowell Observatory (J.~Hall, private communication), and adopting a 
quadratic function for the correlation \citep[cf.][]{hen96}. The details 
of our observations of $\iota$~Hor are listed in Table~\ref{tab1}. Note 
that the uncertainties shown in Table~\ref{tab1} represent only the 
internal errors. The systematic uncertainty from the conversion between 
the CTIO and MWO activity indices is $\sigma_{\rm sys}\sim +0.007$ 
(MWO~$-$~CTIO).

In addition to the single-epoch observation from \cite{hen96}, $S_{\rm 
MWO}=0.225\pm0.005$ on 1992.9479, there are several other Ca~HK 
measurements of $\iota$~Hor in the literature that we can use to probe 
activity variations on longer timescales. \cite{jen06} published a 
recalibration of measurements originally made by \cite{tin02} on 2001.5918 
with $S_{\rm MWO}=0.249\pm0.002$. \cite{gra06} obtained a spectrum on 
2002.9538 with a revised $S_{\rm MWO}=0.226\pm0.01$ (R.~Gray, private 
communication). Finally, \cite{sch09} measured $S_{\rm MWO}=0.246\pm0.03$ 
from a spectrum obtained on 2003.9387. Additional Ca~HK measurements of 
$\iota$~Hor have appeared in the literature, but without sufficient detail 
to determine the precise epoch of the observation and the $S$ index on the 
Mount Wilson scale.


\section{RESULTS\label{SEC3}}

\begin{figure}[t]
  \centerline{\includegraphics[angle=90,width=3.25in]{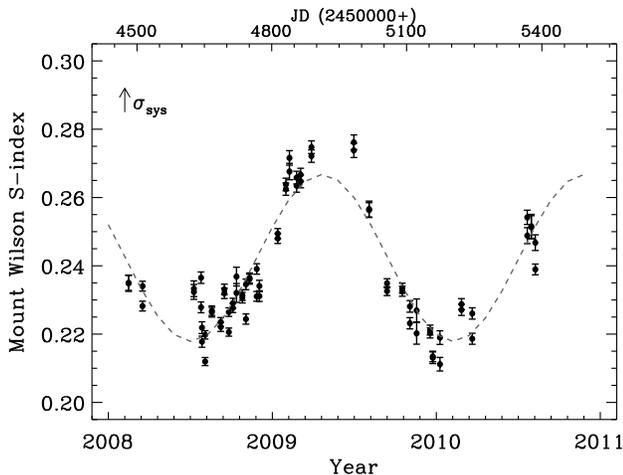}}
  \caption{Chromospheric activity measurements of the F8V star $\iota$~Hor 
  from the southern HK survey \citep{met09}, showing a clear variation with 
  a cycle period of 1.6 years, the shortest cycle measured for a sun-like 
  star. Note that the error bars represent only the measurement errors, and 
  do not include the systematic uncertainty $\sigma_{\rm sys}\sim 0.007$ 
  (arrow).\label{fig1}}
  \end{figure}

Our time-series measurements of the Mount Wilson $S$ index for $\iota$~Hor 
are plotted in Figure~\ref{fig1}. We fit a sinusoid to these data and 
found a cycle period of $P_{\rm cyc} = 1.6$ years around a mean value of 
$\left< S \right> = 0.242$. The variations are not expected to be strictly 
sinusoidal, so the fitted amplitude of $A_{\rm cyc} = 0.024$ does not 
capture the full range of observed values from $S\sim0.21$-0.28, nearly 
30\% of the mean activity level \citep[the solar range is 
$S\sim0.17$-0.20;][]{bal95}. The parameter values of the fit do not change 
significantly when including the few archival data points, and all but the 
\citeauthor{gra06} measurement agree with the extrapolated sinusoid at the 
$1\sigma$ level.

We were initially intrigued that the activity maximum at 2009.3 coincided 
with an epoch of periastron for the eccentric exoplanet in the system 
\citep[cf.][]{nae01}. Although activity induced by hot Jupiters has been 
seen in a few cases \citep[e.g., see][]{wal08}, we would not expect it in 
this case because even at periastron the star-planet separation is 0.7~AU. 
The exoplanet orbital period of 311 days bears no obvious relation to the 
cycle period of 584 days, so this appears to be simply a coincidence.

Asteroseismic observations of $\iota$~Hor by \cite{vau08} in 2006.9 
coincided with the magnetic minimum one cycle prior to the beginning of 
our data set. As first observed in the Sun more than two decades ago 
\citep{lw90}, the global oscillation frequencies are shifted significantly 
from solar minimum to maximum. The amplitude of these frequency shifts was 
predicted to be larger ($\sim$1~$\mu$Hz) for stars hotter than the Sun 
\citep[see][]{met07,kar09}, as was recently confirmed in the F5V star 
HD\,49933 by \cite{gar10}. Future asteroseismic observations near the 
magnetic maxima that coincide with the observing season for $\iota$~Hor in 
late 2010 or 2013 have the best chance of detecting these frequency 
shifts.

\begin{figure}[t]
  \centerline{\includegraphics[angle=90,width=3.25in]{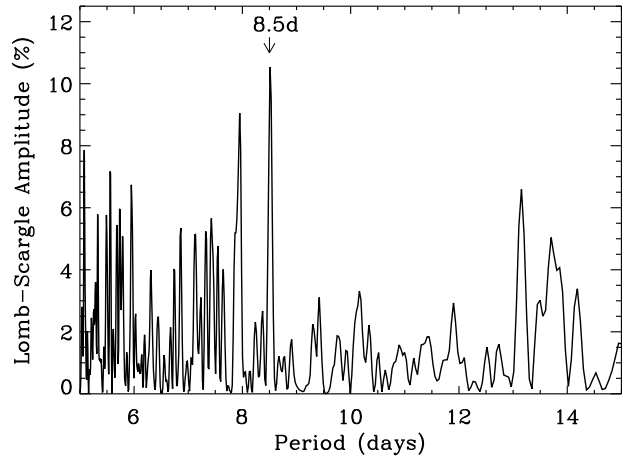}}
  \caption{The Lomb-Scargle periodogram of our Ca~HK measurements after 
  removing the 1.6 year sinusoid, suggesting a rotation period that is
  consistent with Hyades membership for this star.\label{fig2}}
  \end{figure}

The scatter around the sinusoidal variation in Figure~\ref{fig1} is caused 
in part by rotational modulation of individual active regions. Based on 
prior single-epoch measurements of its chromospheric activity level, the 
rotation period of $\iota$~Hor was estimated to be 7.9 days \citep{so97}. 
After normalizing our measurements by the 1.6 year sinusoid, we passed the 
residuals through a Lomb-Scargle periodogram \citep{lom76,sca82} to search 
for the signature of rotation. The results are shown in Figure~\ref{fig2}, 
with the highest peak at $P_{\rm rot}=8.5\pm0.1$ days and a smaller peak 
near 7.9 days. We verified that a single sinusoid with a period of 8 days, 
sampled in the same way as our data, produces a clear and significant peak 
in the periodogram. Of course, the rotational modulation of individual 
active regions will not be coherent over the span of our data set, and 
spots at different latitudes may have different periods. These effects 
will tend to reduce the significance of individual peaks in the 
periodogram. However, a recent analysis by \cite{boi10} of the radial 
velocity measurements from \cite{vau08} found evidence of rotation in 
$\iota$~Hor with periods in the range 7.9 to 8.4 days, consistent with the 
dominant peaks in Figure~\ref{fig2} from our Ca HK measurements. The mean 
rotation period of solar-type stars in the Hyades is $\sim$8.4~days 
\citep{rad95}, so both of these results support the conclusions of 
\cite{mon01} and \cite{vau08} that $\iota$~Hor is an evaporated member of 
the cluster.

If we combine the asteroseismic radius of $\iota$~Hor, $R=1.18\ R_\odot$ 
(S.~Vauclair, private communication) and the rotation period above with 
the measured $v \sin i = 6.47 \pm 0.5$~km\,s$^{-1}$ from \cite{but06}, we 
can obtain an estimate of the inclination angle, $i \sim 60^\circ$. If the 
rotation axis is perpendicular to the orbital plane of the 
exoplanet\footnote{Note that recent measurements of the 
Rossiter-McLaughlin effect for some transiting hot Jupiter systems suggest 
a substantial spin-orbit misalignment \citep[e.g., see][]{sch10}.}, then 
the absolute mass of the planet is only about 15\% larger than the minimum 
mass obtained from the radial velocity orbital solution.


\section{CONCLUSIONS \& DISCUSSION \label{SEC4}}

The immediate question that arises from the observation of such a short 
activity cycle is whether we can understand it in the context of dynamo 
models that are frequently invoked to explain the 11-year solar cycle. It 
is generally difficult to extrapolate solar dynamo models to other stars 
because it is not well understood how the underlying parametrizations 
change for different stellar properties. However, in the case of 
$\iota$~Hor we can take advantage of a lucky coincidence---an F8 star with 
a rotation period near 8 days has a Coriolis number $2\Omega\tau_c$ that 
is very similar to that of the Sun \citep{kr05}. Since the Coriolis number 
characterizes the influence of rotation on convection, it determines to a 
large extent the overall direction and profile of the turbulent Reynolds 
stresses (RS) that drive differential rotation and meridional flow. The 
amplitudes are expected to be different due to the higher luminosity and 
resulting convective velocities (${\rm RS} \sim v_c^2$), but it still 
allows us to make some simple estimates of how various dynamo scenarios 
considered for the Sun are expected to scale to $\iota$~Hor.

From a standard stellar evolution model of $\iota$~Hor with the global 
properties inferred from asteroseismology \citep{vau08}, we deduce that 
the convective velocity $v_c$ is 2.8 times higher than the solar value 
while the relevant timescale $\tau_c$ is 3.2 times shorter (due to a 
smaller pressure scale height). Using simple scaling relations these 
quantities allow us to estimate turbulent diffusivities $\eta_t$, $\nu_t$ 
that are 2.4 times solar. The meridional flow speed $v_m$ is expected to 
be about 2.6 times solar (from angular momentum transport balance, $v_c^2 
\sim v_m\Omega$) and the differential rotation 2-3.3 times solar 
(considering thermal balance $\Omega\Delta\Omega \sim \Delta T$ and 
angular momentum transport balance $v_c^2 \sim \nu_t\Delta\Omega$). 
Additional scaling factors of order unity can arise from the different 
depth and aspect ratio of the convection zone. Nevertheless these 
estimates agree very well with the differential rotation model for an F8 
star presented in \cite{kr05}.

The above properties would lead us to predict a cycle period of 4-5 years 
for an advection dominated flux-transport dynamo ($P_{\rm cyc} \sim 
v_m^{-1}$), or 3.5-5 years for a classical $\alpha$-$\Omega$ dynamo 
($P_{\rm cyc} \sim [\Delta\Omega\alpha]^{-0.5}$, assuming $\alpha\sim 
v_c$). These estimates are both more than a factor of two longer than the 
observed cycle, but for the flux transport dynamo we assumed that the 
underlying conveyor belt would extend to similar latitudes as observed in 
the Sun. Due to the larger aspect ratio of the convection zone in an F8 
star as well as the larger magnetic diffusivity, it is conceivable that 
the flux transport is cut short leading to a significant reduction in the 
dynamo period. Currently, B.~P.~Brown et al.\ (in preparation) are 
exploring dynamo action in solar-type stars rotating at 3-5 times the 
solar rate using global 3D anelastic MHD models. For a sufficiently large 
level of turbulence, cyclic behavior is found with periods of a few years 
or less. These studies demonstrate that computing a detailed 3D model of 
$\iota$~Hor should already be possible.

Prior to the discovery of this 1.6-year magnetic cycle in $\iota$~Hor, the 
shortest measured cycle periods were 2.52 years (HD\,76151) and 2.60 years 
(HD\,190406) from the Mount Wilson survey \citep{bal95}. Both of these 
appeared to be secondary cycles superimposed on a much longer primary 
cycle. \cite{bv07} has suggested, based on the sample of stars with well 
characterized rotation and cycle periods in \cite{sb99}, that the 
``active'' and ``inactive'' branches in the $P_{\rm rot}$-$P_{\rm cyc}$ 
diagram may be caused by two distinct dynamos that are driven in different 
regions of the star. Specifically, \citeauthor{bv07} suggests that the 
active branch may represent a dynamo operating in the near surface shear 
layer, while the inactive branch is driven by the shear layer at the base 
of the convection zone. In this scenario, stars that exhibit cycle periods 
on both branches must have the two types of dynamos operating 
simultaneously. If the 1.6-year cycle in $\iota$~Hor is on the inactive 
branch, then the active branch cycle period is expected to be near 6 
years. Although the previous Ca~HK measurements are sparse, we see no 
evidence of a secular trend in $S_{\rm MWO}$ on longer timescales. 
Continued observations by our program should yield stronger constraints on 
possible slow variations.

Recently, \cite{gar10} detected the signature of a short magnetic activity 
cycle in the F5V star HD\,49933 using asteroseismic measurements from the 
CoRoT satellite \citep{app08,ben09}. Just as in the Sun, where the global 
oscillation modes shift to higher frequencies and lower amplitudes towards 
the maximum of the 11-year solar cycle, HD\,49933 appeared to pass through 
a magnetic minimum during 137 days of continuous observations. An 
additional 60 days of data from an earlier epoch could not place strong 
constraints on the cycle period, but suggested a value between 120 days 
and about 1 year. If confirmed by ground-based monitoring of the Ca~HK 
lines, this would place HD\,49933 (with $P_{\rm rot}=3.4$\,d) in the same 
category of magnetic cycle observed in $\iota$~Hor.

The asteroseismic signature of a short magnetic cycle has also been 
detected in the Sun itself. \cite{fle10} analyzed the low-degree solar 
oscillation frequencies from the BiSON and GOLF experiments and found 
evidence of a quasi-biennial (2 year) signal in both data sets after 
removing the dominant 11-year period. Unlike the 11-year signal, the 
amplitude of the 2-year variation appeared to be largely independent of 
frequency, leading the authors to suggest that the secondary cycle must be 
operating independently. However, the amplitude of the 2-year signal was 
uniformly larger during the maximum of the 11-year cycle, suggesting that 
buoyant magnetic flux might be rising from the base of the convection zone 
and pumping up a near-surface dynamo with the 2-year period. Active branch 
stars with a cycle period of 11 years are expected to show a secondary 
cycle period on the inactive branch around 2 years. However, this normally 
occurs in stars rotating at twice the solar rate, and the identification 
of the two dynamos would then be reversed---with the 11-year dynamo 
operating in the near-surface shear layer, while the 2-year dynamo is 
driven at the base of the convection zone.

If short magnetic activity cycles are common, NASA's {\it Kepler} mission 
should detect them in the asteroseismic measurements of many additional 
stars. In principle such measurements can provide unique constraints on 
the underlying physical mechanism, and {\it Kepler} will also yield 
measurements of some of the key dynamo ingredients. Even without the short 
cadence data for asteroseismology, the high precision time-series 
photometry from {\it Kepler} is sufficient to characterize the surface 
differential rotation through detailed spot modeling \citep[e.g., 
see][]{bas10}. For the brighter asteroseismic targets where the individual 
oscillation frequencies are detectable, the time series will be long 
enough to resolve rotational splitting of the modes into multiplets for 
stars with rotation rates between about 2 and 10 times the solar rate 
\citep{bal08}. Measurements of the rotational splitting as a function of 
radial order can indirectly probe radial differential rotation, since the 
various modes sample slightly different (but overlapping) regions of the 
star. For the very best and brightest asteroseismic targets, {\it Kepler} 
will obtain a frequency precision sufficient to measure the depth of the 
surface convection zone from the oscillatory signal in the so-called 
second frequency differences \citep[e.g., see][]{ver06}. The {\it Kepler} 
mission is expected to document all of these properties in at least a few 
dozen solar-type stars, gradually leading to a broader context for our 
understanding of the solar dynamo.


\acknowledgments 
We would like to thank Fred Walter for scheduling our SMARTS program, 
Manuel Hernandez and Jose Velasquez for conducting the observations at 
CTIO, Jeffrey Hall for assistance in calibrating our chromospheric indices 
to the Mount Wilson scale, and Bill Chaplin for helpful suggestions. The 
southern HK survey began with SMARTS time purchased by High Altitude 
Observatory through Georgia State University, and was supported for two 
years under NOAO long-term program 2008B-0039 with additional time from 
SMARTS partners. The National Center for Atmospheric Research is a 
federally funded research and development center sponsored by the U.S.\ 
National Science Foundation.


\begin{thebibliography}{}

\bibitem[Appourchaux et al.(2008)]{app08} Appourchaux, T., et al.\ 2008, 
\aap, 488, 705

\bibitem[Baglin et al.(2006)]{bag06} Baglin, A., et al.\ 2006, ESA-SP, 
624, 34

\bibitem[Baliunas et al.(1995)]{bal95} Baliunas, S.~L., et al.\ 1995, 
\apj, 438, 269

\bibitem[Ballot et al.(2008)]{bal08} Ballot, J., Appourchaux, T., Toutain, 
T., \& Guittet, M.\ 2008, \aap, 486, 867

\bibitem[Basri et al.(2010)]{bas10} Basri, G., et al.\ 2010, \apjl, 713, 
L155

\bibitem[Benomar et al.(2009)]{ben09} Benomar, O., et al.\ 2009, \aap, 
507, L13 

\bibitem[B{\"o}hm-Vitense(2007)]{bv07} B{\"o}hm-Vitense, E.\ 2007, \apj, 
657, 486

\bibitem[Boisse et al.(2010)]{boi10} Boisse, I., et al.\ 2010, \aap, 
submitted

\bibitem[Borucki et al.(2010)]{bor10} Borucki, W.~J., et al.\ 2010, 
Science, 327, 977

\bibitem[Brown et al.(1989)]{bro89} Brown, T.~M., et al.\ 1989, \apj, 343, 
526

\bibitem[Butler et al.(2006)]{but06} Butler, R.~P., et al.\ 2006, \apj, 
646, 505

\bibitem[Chaplin et al.(2007)]{cha07} Chaplin, W.~J., Elsworth, Y., 
Houdek, G., \& New, R.\ 2007, \mnras, 377, 17

\bibitem[Duncan et al.(1991)]{dun91} Duncan, D.~K., et al.\ 1991, \apjs, 
76, 383

\bibitem[Fletcher et al.(2010)]{fle10} Fletcher, S.~T., et al.\ 2010, 
\apjl, 718, L19

\bibitem[Garc{\'{\i}}a et al.(2010)]{gar10} Garc{\'{\i}}a, R.~A., et al.\ 
2010, Science, 329, 1032

\bibitem[Gray et al.(2006)]{gra06} Gray, R.~O., et al.\ 2006, \aj, 132, 
161

\bibitem[Grundahl et al.(2008)]{gru08} Grundahl, F., et al.\ 2008, IAU 
Symp., 252, 465

\bibitem[Henry et al.(1996)]{hen96} Henry, T.~J., et al.\ 1996, \aj, 111, 
439

\bibitem[Jenkins et al.(2006)]{jen06} Jenkins, J.~S., et al.\ 2006, 
\mnras, 372, 163

\bibitem[Karoff et al.(2009)]{kar09} Karoff, C., et al.\ 2009, \mnras, 
399, 914

\bibitem[K{\"u}ker \& R{\"u}diger(2005)]{kr05} K{\"u}ker, M., \& 
R{\"u}diger, G.\ 2005, \aap, 433, 1023

\bibitem[K{\"u}rster et al.(2000)]{kur00} K{\"u}rster, et al.\ 2000, \aap, 
353, L33

\bibitem[Leighton(1959)]{lei59} Leighton, R.~B.\ 1959, \apj, 130, 366

\bibitem[Libbrecht \& Woodard(1990)]{lw90} Libbrecht, K.~G., \& Woodard, 
M.~F.\ 1990, \nat, 345, 779

\bibitem[Lomb(1976)]{lom76} Lomb, N.~R.\ 1976, \apss, 39, 447

\bibitem[Metcalfe et al.(2007)]{met07} Metcalfe, T.~S., et al.\ 2007, 
\mnras, 379, L16

\bibitem[Metcalfe et al.(2009)]{met09} Metcalfe, T.~S., et al.\ 2009, 
Solar Analogs II workshop, (arXiv:0909.5464)

\bibitem[Montes et al.(2001)]{mon01} Montes, D., et al.\ 2001, \mnras, 
328, 45

\bibitem[Naef et al.(2001)]{nae01} Naef, D., et al.\ 2001, \aap, 375, 205

\bibitem[Radick et al.(1995)]{rad95} Radick, R.~R., et al.\ 1995, \apj, 
452, 332

\bibitem[Rempel(2006)]{rem06} Rempel, M.\ 2006, \apj, 647, 662

\bibitem[Saar \& Brandenburg(1999)]{sb99} Saar, S.~H., \& Brandenburg, A.\ 
1999, \apj, 524, 295

\bibitem[Saar \& Osten(1997)]{so97} Saar, S.~H., \& Osten, R.~A.\ 1997, 
\mnras, 284, 803

\bibitem[Salabert et al.(2004)]{sal04} Salabert, D., et al.\ 2004, \aap, 
413, 1135

\bibitem[Scargle(1982)]{sca82} Scargle, J.~D.\ 1982, \apj, 263, 835

\bibitem[Schlaufman(2010)]{sch10} Schlaufman, K.~C.\ 2010, \apj, 719, 602

\bibitem[Schou et al.(1998)]{sch98} Schou, J., et al.\ 1998, \apj, 505, 
390

\bibitem[Schr{\"o}der et al.(2009)]{sch09} Schr{\"o}der, C., Reiners, A., 
\& Schmitt, J.~H.~M.~M.\ 2009, \aap, 493, 1099

\bibitem[Tinney et al.(2002)]{tin02} Tinney, C.~G., et al.\ 2002, \mnras, 
332, 759

\bibitem[Vauclair et al.(2008)]{vau08} Vauclair, S., et al.\ 2008, \aap, 
482, L5

\bibitem[Verner et al.(2006)]{ver06} Verner, G.~A., Chaplin, W.~J., \& 
Elsworth, Y.\ 2006, \apj, 638, 440

\bibitem[Walker et al.(2003)]{wal03} Walker, G., et al.\ 2003, \pasp, 115, 
1023

\bibitem[Walker et al.(2008)]{wal08} Walker, G.~A.~H., et al.\ 2008, \aap, 
482, 691

\bibitem[Wilson(1978)]{wil78} Wilson, O.~C.\ 1978, \apj, 226, 379

\end{thebibliography}
\end{document}